*Revised manuscript*

**Emergence of Dirac-like bands in the monolayer limit of epitaxial Ge films on Au(111)**

Niels B. M. Schröter, Matthew D. Watson, Liam B. Duffy, Moritz Hoesch, Yulin Chen, Thorsten Hesjedal, and Timur K. Kim*

N.B.M. Schröter, L. B. Duffy, Prof. Y.L. Chen, Prof. T. Hesjedal
Department of Physics and Clarendon Laboratory, University of Oxford
Parks Road, Oxford, OX1 3PU, UK

L. B. Duffy
ISIS, STFC, Rutherford Appleton Lab, Didcot, OX11 0QX, United Kingdom

M.D. Watson, M. Hoesch, T.K. Kim
Diamond Light Source, Harwell Campus, Didcot OX11 0DE, United Kingdom
E-mail: timur.kim@diamond.ac.uk

**Abstract**

After the discovery of Dirac fermions in graphene, it has become a natural question to ask whether it is possible to realize Dirac fermions in other two-dimensional (2D) materials as well. In this work, we report the discovery of multiple Dirac-like electronic bands in ultrathin Ge films grown on Au(111) by angle-resolved photoelectron spectroscopy. By tuning the thickness of the films, we are able to observe the evolution of their electronic structure when passing through the monolayer limit. Our discovery may signify the synthesis of germanene, a 2D honeycomb structure made of Ge, which is a promising platform for exploring exotic topological phenomena and enabling potential applications.

# 1. Introduction

Since the discovery of the extraordinary physical and electronic properties of graphene, there has been an intense effort to synthesize two-dimensional (2D) honeycomb structures based on heavier elements than carbon, in order to realize new topological phenomena that are driven by spin-orbit coupling (SOC), such as the quantum spin-Hall (QSH) [1-3] or quantum anomalous-Hall (QAH) [4, 5] effects. One promising family of materials to host these exotic effects – silicene [6], germanene [6, 7], and stanene [7, 8] – is built out of the group IV elements Si, Ge and Sn, which are predicted to form buckled honeycomb structures. Similar to graphene, these structures are expected to host Dirac fermions with a linear dispersion relation in the vicinity of the $K/K'$ points of their hexagonal Brillouin zones. However, unlike graphene, for which SOC is too small to induce a measureable band gap, silicene, germanene, and stanene are predicted to open a gap at $K/K'$ on the order of ~1.6 meV [6], ~24 meV[6], and ~100 meV [8], respectively, which is crucial to observe and utilize the QSH and QAH effects at elevated temperatures, and may pave the way for future applications. It should be noted that Dirac fermions can also be realized in other 2D materials, such as binary honeycomb structures [9] or non-symmorphic 2D materials [10].

For the cases of silicene and stanene, a large number of experimental studies investigated the electronic structure of ultra-thin films of Si and Sn on various metallic and insulating substrates [11-15]. Although initial reports claimed the existence of Dirac dispersions for silicene on Ag(111) [11], more recent studies found that these are likely to be caused by substrate interactions [16-20].

In contrast to its Si- and Sn-based cousins, only very few experiments have so far investigated the electronic structure of germanene. Earlier studies reported its synthesis on the metallic substrates Pt(111) [21], Al(111) [22], Au(111) [23], as well as on $Ge_2Pt$ [24] and $MoS_2$ [25].

Scanning tunneling spectroscopy (STS) studies found evidence for a Dirac dispersion in some of these systems [14], but due to the lack of momentum resolution, it was not possible to disentangle the signal from the Dirac fermions and other bands in the vicinity of the Fermi level. An angle-resolved photoelectron spectroscopy (ARPES) study on a thick film of Ge grown on Au(111), corresponding to about ~4.5 monolayers (ML) of germanene, reported the appearance of a quasi-linear band that crosses the Fermi level close to the point of the underlying Au(111) substrate surface Brillouin zone [26]. However, further investigations are required to determine how this additional band is related to the predicted Dirac fermions in single-layer germanene, or whether it may be induced by the underlying metallic substrate, as was suggested for similar bands found in silicene. For the growth of germanene on Al(111), a buckled structure with a relatively simple registry between 2x2 monolayer Ge on a 3x3 Al(111) substrate was shown by Derivaz *et al.* [22]. However, *ab-initio* calculations have suggested that a strong hybridisation between the Ge and the Al metallic substrate bands would wash out any sign of the Dirac dispersions [27]. On the other hand, some *ab-initio* calculations would suggest that Ge on Ag or Au(111) could be a more promising route to realise Dirac dispersions [27, 28], while others suggest that interactions with the substrate Au d bands will destroy germanene's Dirac dispersion for Ge on Au(111) [29]. Thus, the presence of Dirac fermions in graphene-analogues supported by metallic substrates requires experimental confirmation.

In the present work, by performing comprehensive ARPES measurements, we study the electronic structure of ultra-thin Ge films grown on Au(111). By tuning the thickness of the Ge layer, we are able to track the evolution of the resulting band structure from a sub-ML to the trilayer regime, which allows us to identify bands that are replicated from the Au(111) substrate. In the ML limit, by tuning the incident photon polarization, we are able to unmask a number of previously unreported linearly dispersing Dirac-like bands, which may originate from rotationally disordered germanene.

## 2. Results and discussion
### 2.1. Film characterization

To experimentally determine the thickness of the Ge films, we exploited a shift in the binding energy of the Ge $3d$ doublet peak for different chemical environments of the Ge atoms. As shown in Figure 1(a), for a Ge film of nominal 1.0 Å thickness (upper panel), the photoemission intensity is dominated by the Ge $3d$ doublet at higher binding energies, similar to previous reports [15]. When substantially increasing the film thickness to 7.2 Å (lower panel), the photoemission intensity, which is extremely surface-sensitive, is instead dominated by a doublet at lower binding energies, since the chemical environment of Ge in an epitaxial multilayer sample is distinct from the sub-ML case of the Ge-Au interface. At a nominal deposition of 2.6 Å, the two doublets are observed to coexist, supporting that this sample is close to the ML limit. We furthermore performed scanning tunneling microscopy measurements to confirm that a Ge monolayer is approximately 2.5 Å thick, which can be found in the supplementary material.

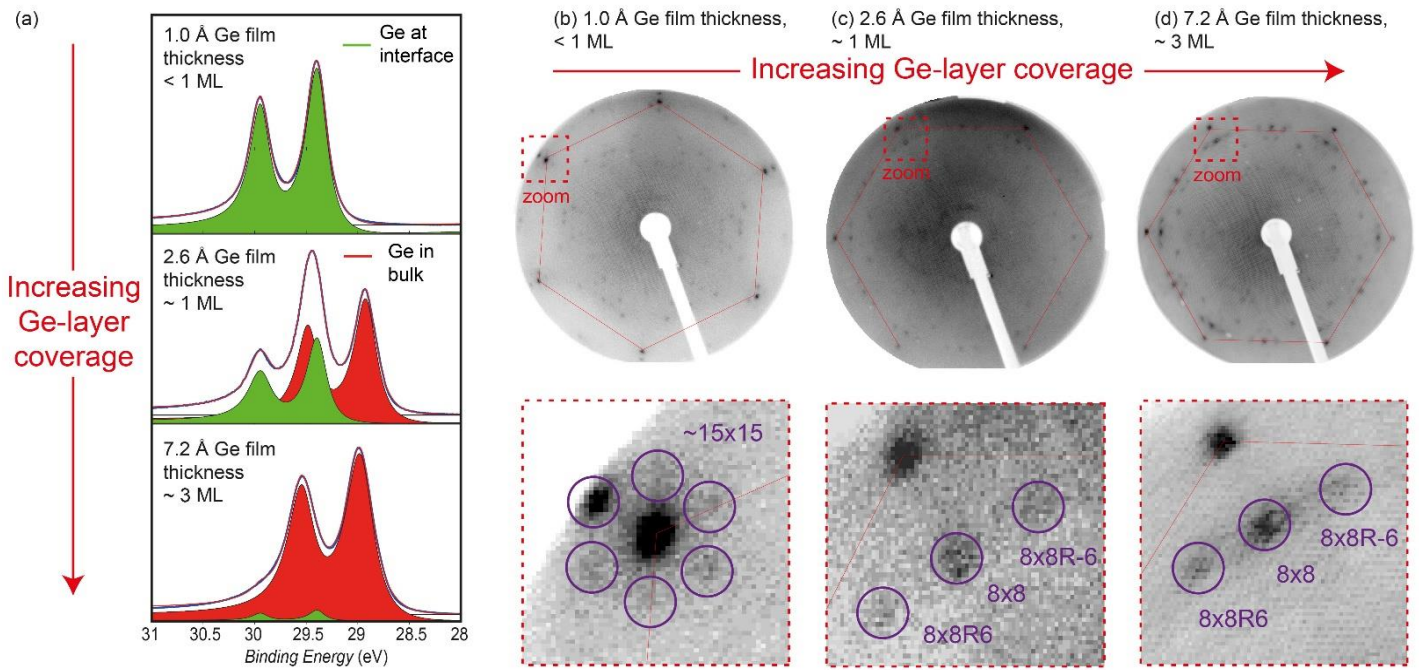

**Figure 1. Film characterization by core level spectroscopy and LEED**. **(a)** Core level spectroscopy of the Ge $3d$ electrons, taken with a photon energy of 120 eV, for varying film thickness. The green and red shaded areas indicate the fitting of interface and bulk peaks, respectively. **(b-d)** Low energy electron diffraction (LEED) images for 1.0 Å, 2.6 Å and 7.2 Å nominal Ge coverage, respectively, corresponding to sub-monolayer (ML), ML, and trilayer

regime. Upper panels show the full LEED image, where the red hexagon indicates the first order diffraction spots from the Au(111) substrate. The lower panel shows a close-up of the superstructure diffraction peaks. The rotation of primary Au(111) LEED spots (red hexagon) between (b) and (c-d) is due to the use of a different but equivalent Au crystal for the sub-monolayer coverage.

The formation of a ML of Ge is also evident as a structural transition in the low-energy electron diffraction (LEED) images shown in Figure 1(b-d). At sub-ML fractional coverage, the most prominent peaks lie in a regular hexagon around the Au substrate peaks. This indicates that the sub-ML Ge coverage forms a long-range superstructure, which can be approximately indexed to a 15x15 reconstruction (of the Au(111) surface lattice, see supplementary material (SM) for more details). Upon further Ge deposition, the LEED pattern changes to an 8x8 superstructure once the first ML is completed, as shown in Figure 1(c). Davila *et al.* proposed a possible structural model for this phase with a registry between a germanene cell and a 8x8 supercell of Au(111) [26]. However, not all LEED spots fit the 8x8 model, as further discussed in the SM. This 8x8 reconstruction is always observed in samples with a thickness from ~1 to at least 3 MLs (as can be seen in Figure 1(d)), indicating epitaxial growth in this regime and no further structural phase transitions.

## 2.2. Band structure evolution with film thickness

In Figure 2 we present the evolution of the Fermi surface with film thickness, as measured by ARPES. We focus on the area around the $\overline{K}$ point of the Au(111) surface Brillouin zone (BZ), where the strongest changes in the electronic structure are observed. For pristine Au(111), the projections of the 3D Au Fermi surface onto the Au(111) surface BZ form a quasi-triangular projected band gap around the $\overline{K}$ point. This can be seen from the experimental ARPES data shown in Figure 2(a) and 2(e), as well as our *ab-initio* calculation of the Au band structure, shown as solid red and blue lines in Figure 2(i). When depositing a sub-ML Ge film (nominal thickness 1.0 Å), large triangles appear around the $\overline{K}$ point, shown in Figure 2(b) and 2(f). They can be understood as replicas of the original Au bulk bands, shown as red and blue dashed lines in Figure 2(j). The surface wavevector by which the substrate bands are shifted corresponds to the ~15x15 superstructure diffraction peaks observed in the LEED measurement shown in Figure 1(b). When increasing the Ge film thickness to the ML limit (approximately 2.6 Å), a smaller triangular Fermi surface emerges, as can be seen in Figure 2(c) and 2(g), and which is illustrated by the solid green line in Figure 2(k). Simultaneously, the intensity of substrate band replica is decreased compared with the sub-ML Ge coverage. When increasing the Ge layer thickness further to 7.2 Å, the small triangle remains, while the substrate induced replica bands are now fully suppressed. At first sight, the small triangle could be explained by another replica of the substrate bands, similar to the ~15x15 reconstruction which appears at 1.0 Å nominal coverage, but with a larger shifting vector, as is illustrated by the dashed lines in Figure 2(l). However, as we will show below, there is strong evidence that these bands may actually originate from the Ge layers.

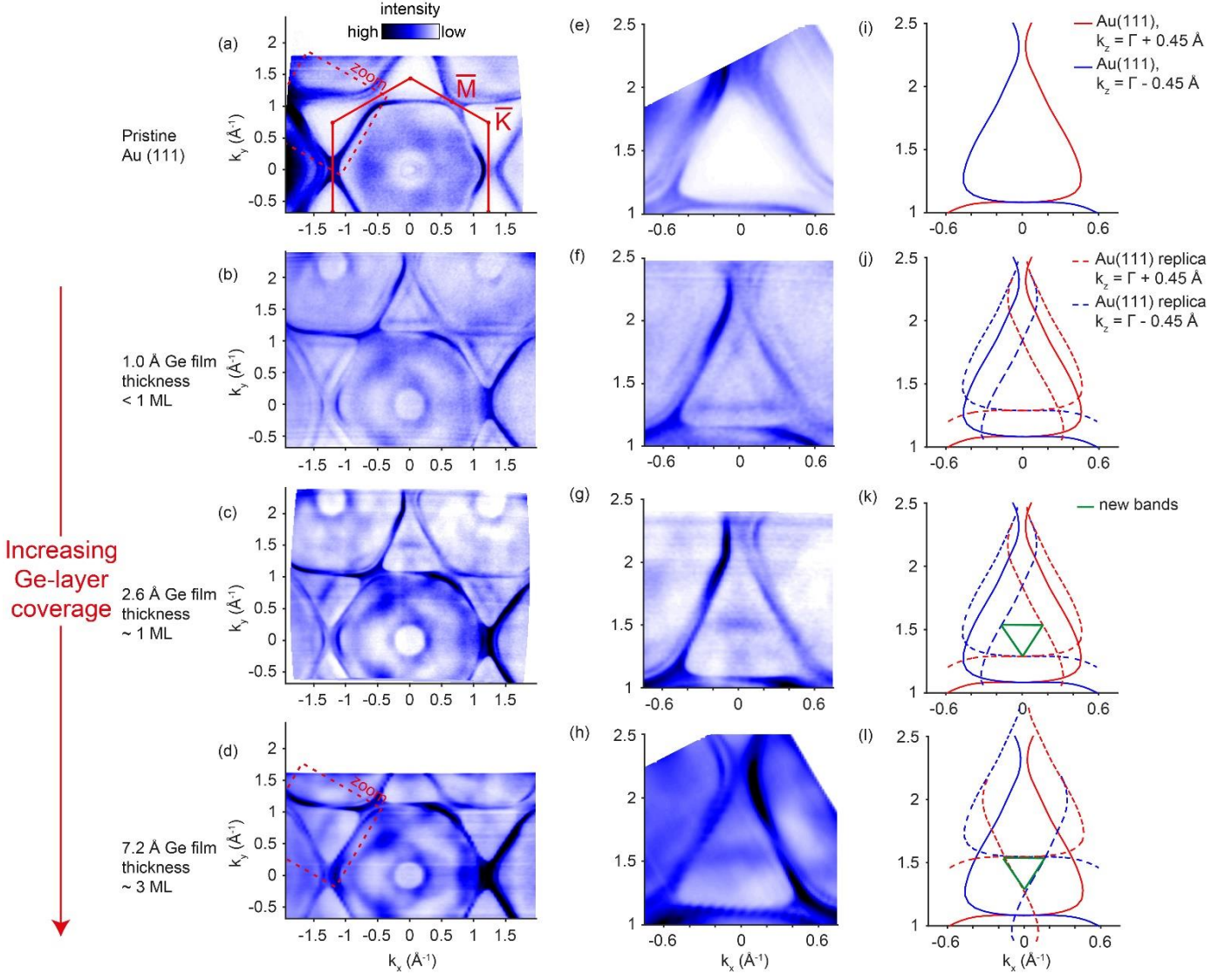

**Figure 2. Fermi surface evolution with Ge film thickness**. **(a -d)** Large scale Fermi surfaces for pristine Au(111), and 1.0 Å, 2.6 Å, and 7.2 Å Ge film thickness. The red solid lines show the surface BZ of the Au(111) substrate, while the red dashed lines indicate the zoomed in regions in (e) and (h). **(e-h)** Fermi surface close to the quasi-triangular band gap at the $\overline{K}$-point of the Au(111) substrate, for pristine Au(111), 1.0 Å, 2.6 Å, and 7.2 Å Ge film thickness, respectively. **(i)** *Ab-initio* calculation of the Au(111) substrate Fermi surface, projected on the (111) surface, at $k_z = \Gamma + 0.45$ Å$^{-1}$ (red solid line) and $k_z = \Gamma - 0.45$ Å$^{-1}$ (blue solid line). **(j)** As in (i), however, with additional replica bands of the Au(111) substrate (red and blue dashed lines), forming a triangle. **(k)** As in (j), however, showing additional bands (see green triangle). **(l)** As in (k), however, the Au(111) replica bands (red and blue dashed lines) are now shifted by a larger reciprocal lattice vector.

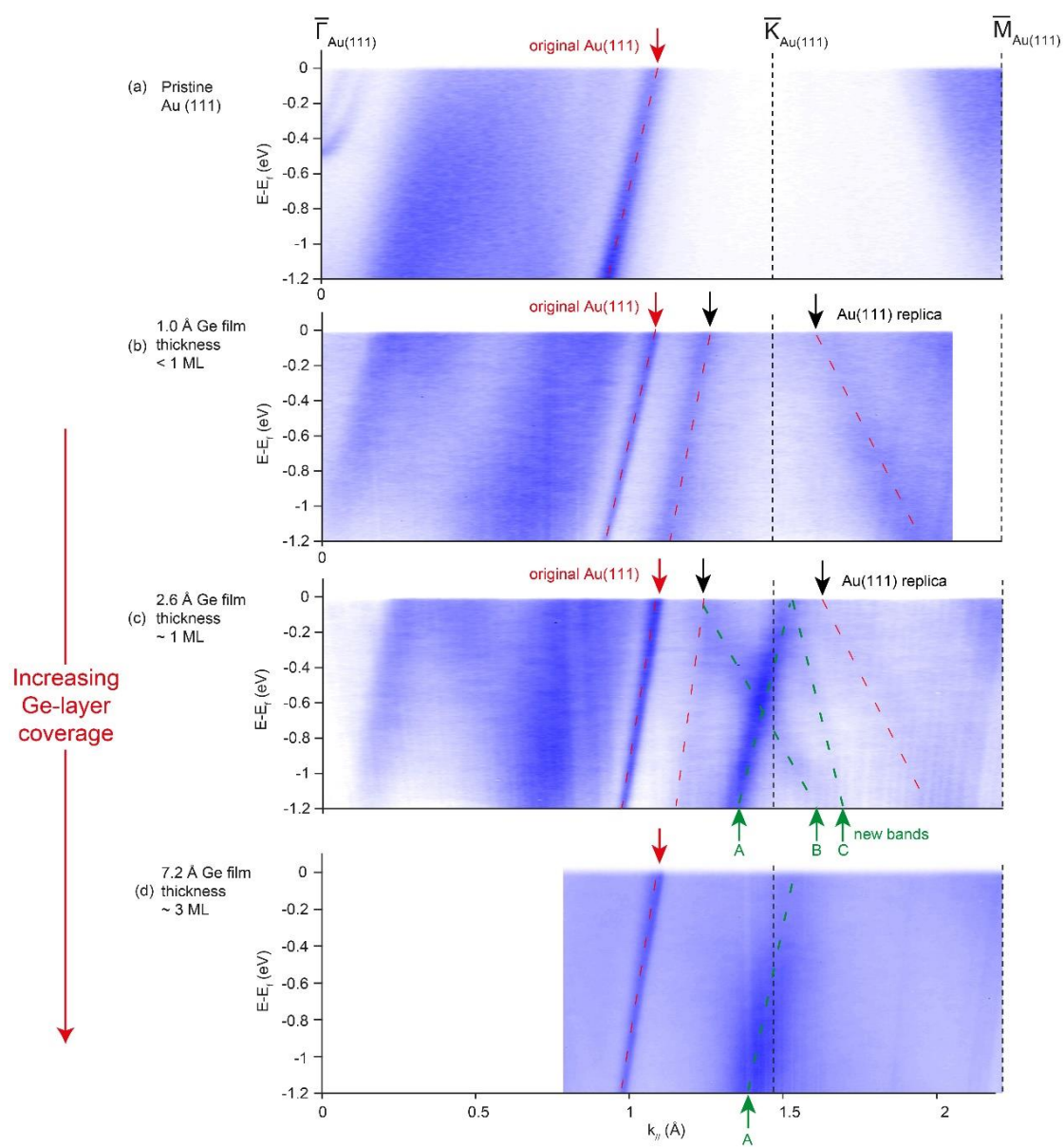

**Figure 3. Energy-momentum dispersion evolution with nominal Ge film thickness**. Measurements taken with linear horizontal (LH) light polarization at a photon energy of 120 eV. **(a-d)** Dispersions for pristine Au(111), and 1.0 Å, 2.6 Å, and 7.2 Å Ge film thickness, respectively. The red arrows indicate a strong Au(111) substrate band. The black arrows indicate replica bands from this substrate bands. Original and replica bands of Au(111) are indicated by red dashed lines as guide to the eye. The green arrows (A, B, C) indicate the new bands, which form when a nominal ML coverage is reached. They are indicated by green lines, serving as guide to the eye.

The drastic changes of the electronic structure as a function of Ge layer thickness can also be observed in the energy-momentum dispersions along the $\overline{\Gamma} - \overline{K} - \overline{M}$ direction of the Au(111) surface BZ, which are shown in Figure 3. For reference, the measurement of the pristine Au(111) surface is shown in Figure 3a, which reveals a Rashba surface state at $\overline{\Gamma}$ and a strong Au *sp* band dispersing along $\overline{\Gamma} - \overline{K}$, indicated by a red arrow. When depositing a sub-ML Ge film (nominal thickness 1.0 Å), a suppression of the surface state is observed in Figure 3(b), and two new bands appear, which are indicated by the black arrows. The new bands along $\overline{\Gamma} - \overline{K}$ are generated by shifting the Au(111) substrate band along the $\overline{\Gamma} - \overline{K}$ direction, whilst the new band along $\overline{K} - \overline{M}$ is replicated from another Au(111) substrate band in the next BZ, which itself does not pass through $\overline{K} - \overline{M}$. These replica bands are responsible for the formation of the triangular Fermi surface, as shown in Figure 2(b) and 2(f). When increasing the nominal Ge layer thickness to 2.6 Å, as shown in Figure 3(c), three new bands appear (green arrows A, B, C), two of which (A, B) form the small green triangles shown in Figure 2(k), while band C becomes more pronounced when switching the polarization from linear-horizontal to linear-vertical, which will be discussed in the next section. Increasing the nominal film thickness to 7.2 Å leads to a broadening of these new bands, which can be seen in Figure 3(d), as well as the disappearance of the replica bands from Figure 3(b).

## 2.3. Emergence of multiple Dirac-like bands in the monolayer limit

In Figure 4 we present evidence that the new bands emerging at the ML-limit may originate from the formation of germanene, rather than the metallic substrate. By switching the polarization of the incident photons with respect to the sample plane, we can unmask a set of bands that were previously hidden due to photoemission selection rules [30]. These selection rules are determined by the relationship between the symmetry of the involved electron orbitals and the symmetry of the incoming light, and can lead to the suppression of photoemission intensity from certain bands. To make these bands visible, the selection rules can be relaxed by changing the symmetry of the incoming light, i.e. changing its polarization. For the present case, we changed the polarization from linear-horizontal to linear-vertical, which revealed several previously unreported bands. Those bands are shown in Figure 4(a) and 4(b) for the Ge films with 2.6 Å and 7.2 Å nominal thickness, respectively. They form a pair of Dirac-like dispersions with apexes centered at the $\overline{M}$ point and slightly off the $\overline{K}$ point of the Au(111) surface BZ. Although previous *ab-initio* calculations concluded that the strong hybridization between the substrate bands and the germanene bands will destroy any linear bands close to the Fermi level [16], the 8x8 superstructure observed in the LEED image in Figure 1(c) may provide a loophole for germanene's Dirac cones to survive by folding them into the projected band gap of the substrate band structure. The possible folding mechanism is illustrated in Figure 4(c), which shows the surface BZ of the Au(111) substrate (red lines and letters), germanene (green lines and letters, based on lattice constant $a_{germanene}$ = 4.4 Å, which is ~10% larger than the theoretical value of freestanding germanene [31]), and the 8x8 superstructure BZ (blue lines). The blue arrows indicate the folding vectors of the 8x8 superstructure, which connect the *K* point of the germanene BZ with the $\overline{M}$ point as well as a point slightly off the $\overline{K}$ point of the substrate BZ. These positions are in excellent agreement with the location of the apexes of the Dirac-like dispersions in our ARPES measurements. The Fermi surface of the 2.6 Å film is shown Figure 4(d) and shows two almost parallel bands. This deviation from the expected conical shape of a Dirac dispersion may be caused by slight rotational disorder, similar to what has been reported for rotationally disordered Graphene on copper [32]. We simulated the effect of rotational disorder on a circular Dirac cone, which can be found in the supplementary material, and nicely reproduces the parallel bands shown in Figure 4(d). A momentum-distribution curve (MDC) analysis of the Dirac-like bands close to the point of the Au(111) surface BZ in the 2.6 Å thick Ge film, shown in Figure 4(e), reveals that the extrapolated apex of the cone is located at 90 meV binding energy. The Fermi velocities of the bands are ~$10^6$ m/s, which is similar to the order of magnitude reported in graphene [33].

We would like to note that, although there are currently no *ab-initio* calculations for germanene/Au(111) that include the large experimentally observed 8x8 superstructure, the folding of germanene's Dirac cone to the Γ-point of the reconstructed BZ has recently been proposed for smaller superstructures based on *ab-initio* calculations [27]. Because of the folding, a strong hybridization with the substrate bands was avoided and the Dirac dispersion was preserved. It is therefore a plausible assumption that a similar mechanism can also exist for the larger 8x8 reconstruction, which needs to be confirmed by future theoretical work.

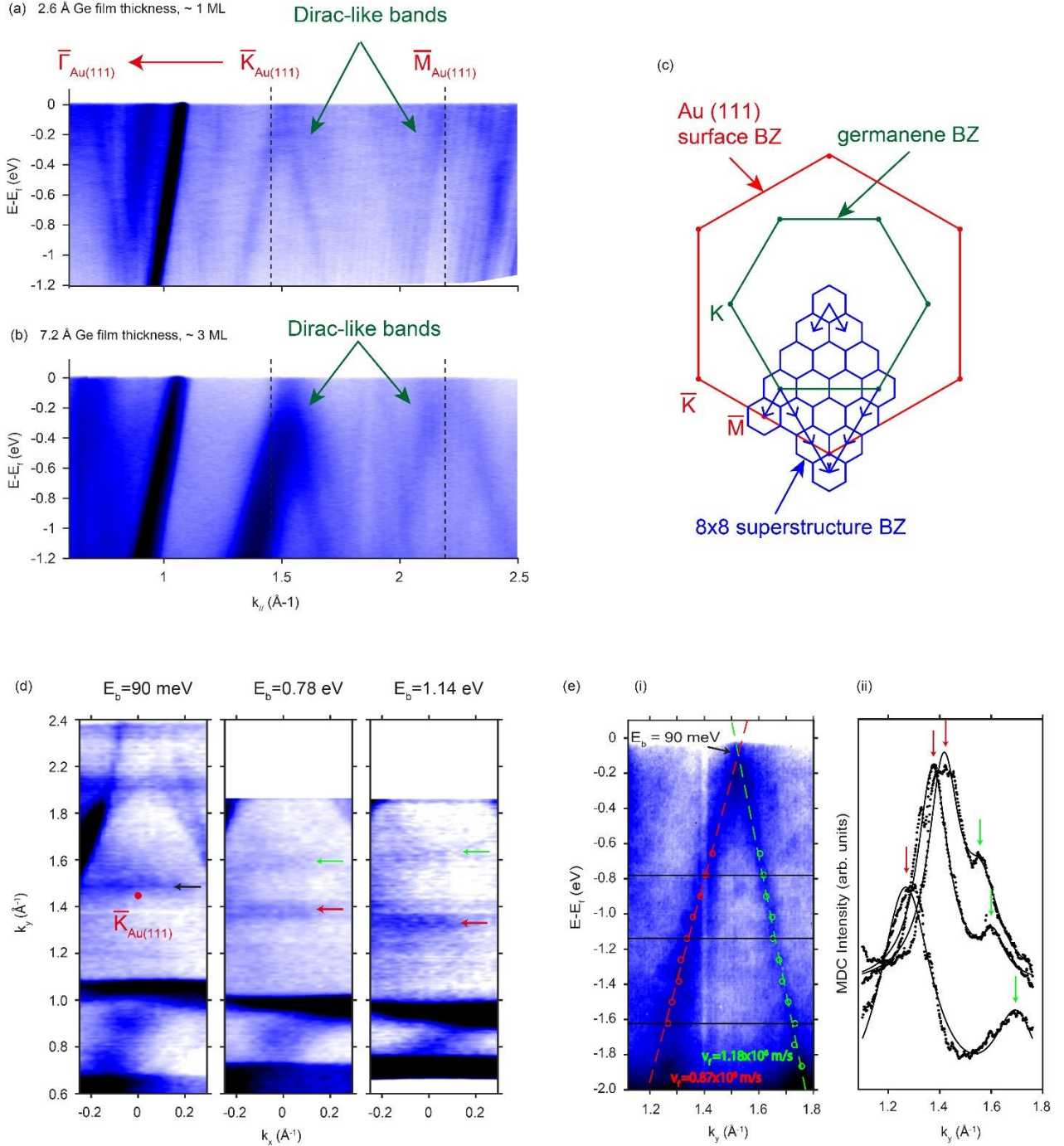

**Figure 4: Multiple Dirac-like cones emerging close to high symmetry points of the Au(111) surface BZ.** All measurements were taken with linear horizontal (LV) light polarization at a photon energy of 120 eV. **(a,b)** Energy-momentum dispersions for 2.6 Å and 7.2 Å nominal film thicknesses. Red arrows indicate Dirac-like bands. **(c)** Illustration of the Au(111) surface BZ (red lines), theoretically predicted germanene BZ (green line), and 8x8 superstructure BZ (blue lines). **(d)** Equal energy surfaces of the nominally 2.6 Å thick Ge film, close to the point of the Au(111) surface BZ, for binding energies $E_b$=0.09 eV, $E_b$= 0.78 eV, $E_b$= 1.14 eV. The black arrow indicates the apex of the Dirac-like cone, while the green and red arrow indicates its two legs. The red dot indicates the $\overline{K}$ point of the substrate BZ. **(e,i)** Detail of the Dirac-like dispersion in nominally 2.6 Å thick Ge film. The red and green points indicate the peaks of the momentum-distribution curve fitting (MDC). The black lines indicate the binding energies for the MDC curves shown in (e,ii). **(e,ii)** MDCs for the black lines shown in (e,i). The red and green arrows indicate the peaks of the fitting curves.

## 3. Conclusion

In summary, we have reported detailed core-level, LEED, and ARPES measurements of ultra-thin Ge films grown on Au(111), with nominal thicknesses between 1.0 Å and 7.2 Å. Our ARPES spectra reveal Au(111) replica bands for the thinnest films, and the emergence of Dirac-like dispersions at the ML limit, which persists for thicker films up to at least 7.2 Å. These bands may be caused by the folding of germanene's Dirac cones to other positions in momentum-space, due to the presence of an 8x8 surface superstructure, which prevents their hybridization with the substrate bands, similar to what has been proposed by recent ab-initio calculations for smaller reconstructions [27]. This is a remarkable finding since the interaction with metallic substrates was previously considered to preclude the existence of Dirac fermions in germanene on Au(111). Our results provide a strong motivation for future experimental studies investigating the precise microscopic structure of ultra-thin Ge films on Au(111), for instance by grazing incidence XRD. Such a structural determination will be necessary as an input for ab-initio band structure calculations that can consider the large 8x8 superstructure that we found experimentally. It will be furthermore interesting to probe the unoccupied states of the Dirac-like bands to observe the upper half of the Dirac cone that was proposed in the present work, which may be achieved via electron doping or pump-probe ARPES experiments. Moreover, it will be interesting to try to reduce the influence of the Au(111) substrate on the band structure of the Ge thing films, e.g. by decoupling via hydrogen intercalation.

## 4. Methods

The ARPES measurements were performed on beamline I05 (Diamond Light Source, UK) [34]. The Au(111) substrate and Ge films were prepared and grown *in-situ*. The layer thickness was determined via the calibration of the deposition rate with a quartz crystal microbalance, and the observation of core level chemical shifts. Before the deposition of the Ge layers, a Bi layer of 0.8 Å nominal thickness was deposited as a surfactant. ARPES measurements were taken with a photon energy of 120 eV with linear horizontal and vertical polarization. The photoelectron energy and angular distributions were analyzed with a SCIENTA R4000 hemispherical analyzer. The measurement temperature was 10 K and the sample remained in a vacuum of <$5\times10^{-10}$ Torr throughout the measurements. The angular resolution was 0.2°, and the overall energy resolution was better than 25 meV. The *ab-initio* calculations were performed using the generalized gradient approximation and the full-potential linear augmented plane-wave basis within the Wien2k package [35].

**Acknowledgements**

We thank Alexander A. Baker (formerly Oxford Physics, now Lawrence Livermore Labs) for help with the growth of the initial Ge films. We thank Katie Winter and Qixin Yang for their support during the STM measurements. We thank Luke C. Rhodes for his support during the ARPES beamtimes. We thank Diamond Light Source for access to beamline I05 (proposal number SI12799) that contributed to the results presented here. N.B.M.S. acknowledges the support by Studienstiftung des deutschen Volkes. Y.L.C. and T.H. acknowledge the support of the EPSRC Platform Grant (Grant No EP/M020517/1)

# Supplementary material

**Emergence of Dirac-like bands in the monolayer limit of epitaxial Ge films on Au(111)**

Niels. B. M. Schröter, Matthew D. Watson, Liam B. Duffy, Moritz Hoesch, Yulin Chen,

Thorsten Hesjedal, and Timur K. Kim

# 1. Au 4f peaks: suppression of Au(111) surface state and formation of new interface environment

The 4*f* core levels of pristine Au(111) surfaces display splittings within the *$4f_{7/2}$* and *$4f_{5/2}$* peaks due to the different chemical shifts in the bulk and on the surface, as can be seen in Figure S1. However, the Au(111) surface peak is completely suppressed by the time the first ML of Ge is deposited, as a new Au-Ge interface environment develops. This interface layer has its own signature (labelled Int) as a shoulder at higher binding energy than the bulk peaks. The overall intensity of the Au substrate core levels is reduced as additional Ge is deposited on top, but the interface layer becomes relatively more pronounced compared to the bulk contribution as the layers build up. These results are comparable to the data shown in M. E. Dávila, L. Xian, S. Cahangirov, A. Rubio, G. L. Lay, New J. Phys. 2014, 16, 095002.

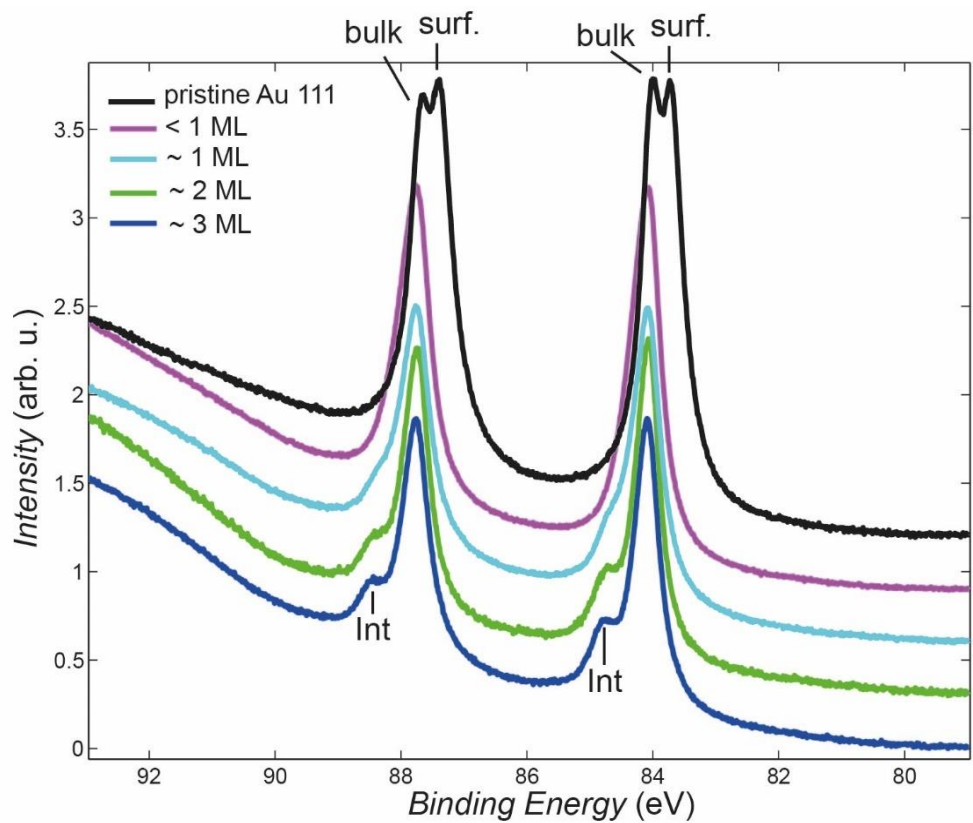

**Figure S1:** Stacked XPS spectra covering the Au 4*f* core levels, obtained with 120 eV photon energy at 10 K in normal emission. Intensities are normalized to background at 80 eV binding energy.

## 2. The role of Bi as a surfactant

The role of a surfactant in thin-film growth is to encourage a change in the growth mode from 3D to epitaxial 2D growth. Sb (isovalent to Bi) was shown to be an effective surfactant for Ge film growth on Si(111) (T. Schmidt, R. Kröger, T. Clausen, J. Falta, A. Janzen, M. Kammler, P. Kury, P. Zahl, M. Horn-von Hoegen, Appl. Phys. Lett. 2005, 86, 111910), thus Bi is likely to play this role in germanene growth as well.

**Bi 4f core levels: Bi rises to the top**

A well-calibrated small quantity of Bi surfactant is deposited first onto the Au substrate, before the Ge deposition. However, it seems that after the growth, Bi is always found on the top surface of the sample, once the few-layer germanene is grown. The evidence for this is that the Ge $4f$ peak height is completely independent of germanene thickness, as shown in Figure S2. This is in contrast to the suppression of the underlying substrate Au $4f$ peaks, which occurs due to the surface-sensitivity of photoemission.

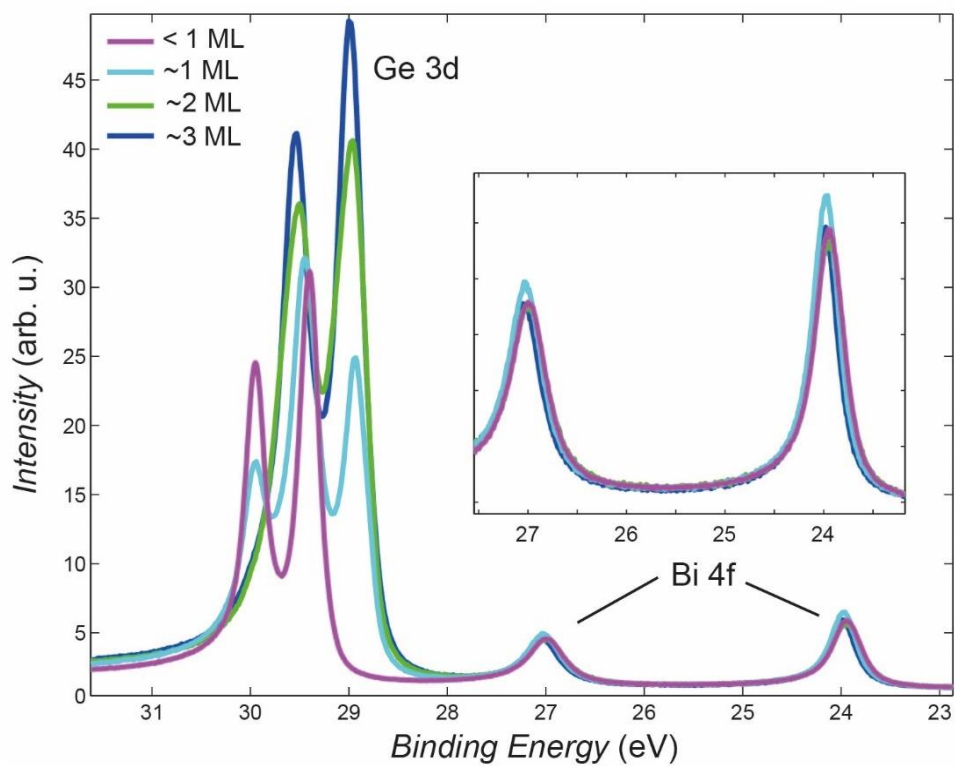

**Figure S2:** Core level photoemission spectrum containing Bi 4*f* and Ge 3*d* peaks. The intensity of the Bi 4*f* peaks is independent of thickness. Data obtained with 120 eV photon energy. The data is normalized to the background.

## 3. Film characterization: LEED

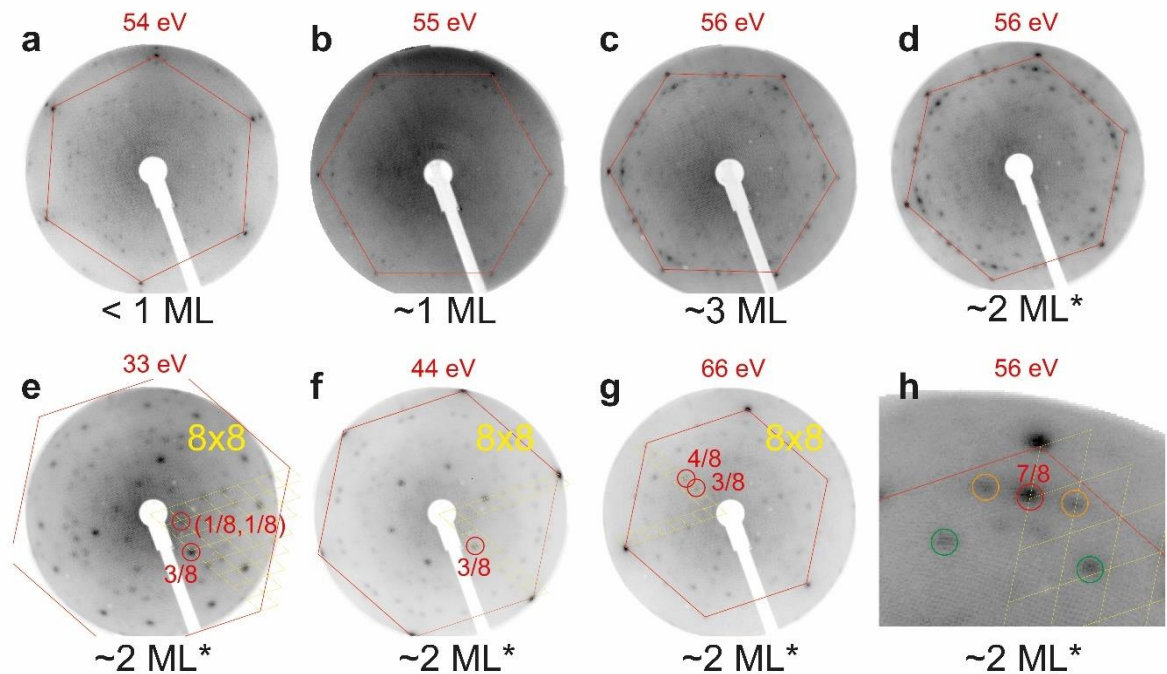

**Figure S3:** a) LEED obtained at ∼55 eV incident energy, on different samples. Red hexagons are used to indicate the first order Au(111) substrate peaks. Note that due to small distortions in the LEED images, in some places the spots deviate slightly from a perfect hexagon. Apart from the sub-ML coverage, all subsequent layers display qualitatively equivalent spectra, although the quality of the spectra is sample dependent (for a mixture of intrinsic and extrinsic reasons). The 2 ML* sample was not used for ARPES measurements. e-g) LEED at various energies on the 2 ML* sample; some prominent peaks associated with the dominant 8×8 reconstruction are circled in red. h) Detail of the 56 eV data from panel d). Orange and green circles highlight spots which do not conform to the 8×8 reconstruction, as discussed.

**Submonolayer case: linking the LEED pattern with replica bands observed by ARPES**

For the sub-ML case, the LEED pattern is qualitatively different to both the pristine Au(111) and the epitaxial few-layer films. One study (M. E. Dávila, G. Le Lay, Sci. Rep. 2016, 6, 20714) indexed a sub-ML LEED pattern of Ge on Au(111) to a combination of 19×19R(23.4°), 5×5, and 7×7R(19.1°) reconstructions, suggesting a number of variants are possible depending on the coverage; here the inclusion of Bi is also likely to play a role. In our sub-ML sample, a few weaker spots in the LEED pattern might be associated with a 5×5 pattern, but the most prominent spots are found as a regular hexagon around the first order substrate peak, which could be indexed to approximately a 15×15 surface reconstruction. Due to this surface superstructure, it is to be expected that replica substrate bands could be observed in ARPES measurements, with the magnitude of the shift in momentum space being linked to the periodicity of the of the superstructure, as is discussed in the main text.

**LEED patterns in the few-multilayer regime**

In Figure S3b-d, we present LEED patterns at different coverages in the range of 1-3 ML, which show no qualitative change in the LEED pattern. This would appear to indicate that, at least in this regime, germanene grows epitaxially. This pattern was also seen by Davila et. al. (M. E. Dávila, G. Le Lay, Sci. Rep. 2016, 6, 20714), who suggested an 8×8 reconstruction of germanene on Au(111) (i.e., 8×8 relative to the Au(111) lattice). Indeed, we find that the peaks with the strongest intensity do match well to an 8×8 reconstruction, as indicated in Figure S3e-h. However, there are also indications that there are features beyond the 8×8 reconstruction (which can also be weakly observed in the LEED pattern of Davila *et. al.* (M. E. Dávila, G. Le Lay, Sci. Rep. 2016, 6, 20714)). Three different sets of peaks do not obey the 8×8 reconstruction. First, as seen most clearly in Figure S3d, the peak pattern seen around the main Au(111) substrate peaks is replicated at a 30° rotation, indicating the likely presence of 8×8R30° domains. Second, the spots marked in orange in Figure S3h, do not conform to any allowed 8×8 diffraction peak, but instead they may be attributed to a 6° rotation of the 8×8 structure (i.e., 8×8R6°). It seems probable that the azimuthal orientation of the Ge film is not strongly constrained, which may be related to the large size of the reconstruction. Third, the peaks marked in green in Figure S3h do not conform to the 8×8 structure. These may be tentatively associated with Bi surface ordering due to the fact that these peaks alone disappear in the range of 250-300°C (see below), as well as the fact that these peaks do not appear in the LEED data of Davila *et. al.* (M. E. Dávila, G. Le Lay, Sci. Rep. 2016, 6, 20714) where no surfactant was used.

**Annealing analysis of a 3 ML sample**

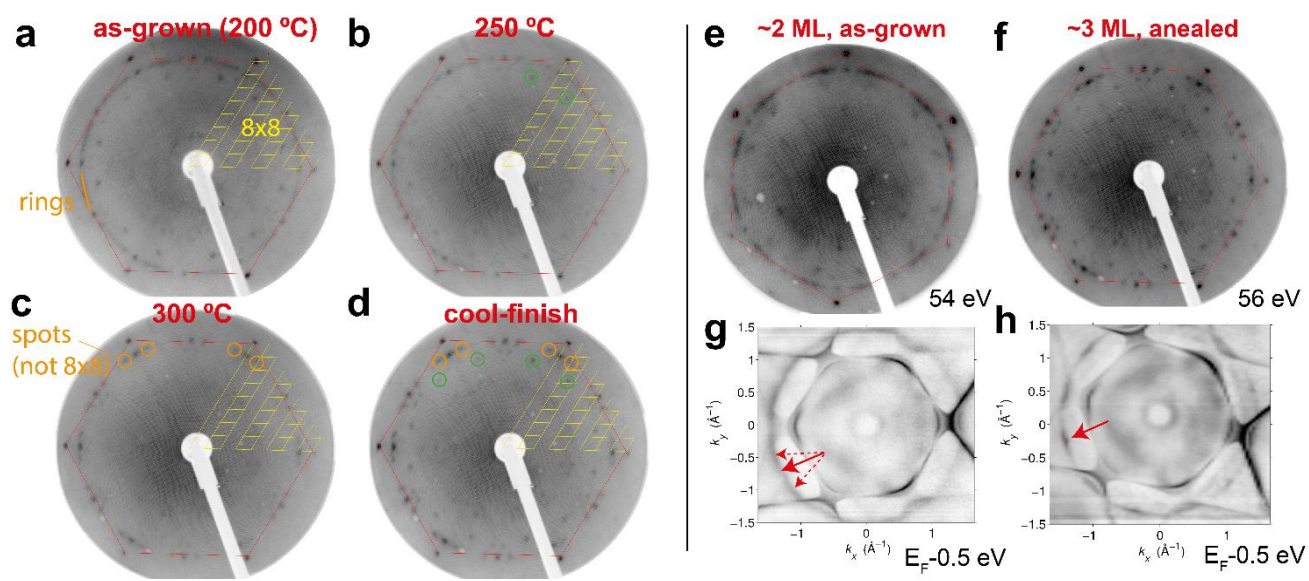

**Figure S4:** Effect of annealing on a 3 ML germanene film.

The as-grown 3 ML film features some spots consistent with an 8x8 reconstruction, but in place of the 8×8R6° peaks, a ring-like continuum of intensity from 6° to 24° rotation. It is likely that some rotational disorder of the growth occurs as the film thickness increases, which manifests as a ring-like structure. After annealing at 300°C, as shown in in Figure S4c, these rings disappear and instead 8×8R6° spots emerge. This indicates that the annealing process aids the azimuthal rotational ordering of the germanene structure on Au(111), though there may still be several preferred orientations. In addition, the main 8×8 peaks appear sharper after the annealing process, indicating that the film quality is increased by the annealing.

Additionally, a set of spots labelled with green rings disappear at temperatures between 250 and 300°C reversibly, since they are present again in the final cooled data. One possible explanation for this is that these peaks are associated with the Bi atoms which are likely to sit in an ordered arrangement on the top surface of the sample. We also note that this annealing procedure gave no significant difference to the XPS spectrum, i.e., no significant desorption occurred at these temperatures.

Annealing also influences the quality of our ARPES data, as can be seen in Figures S4e-h. While the ~2ML as-grown film without annealing shows to rotationally smeared LEED spots (Figure 4e) and rotationally smeared Dirac bands (indicated by red arrows in Figure 4g), the ~3ML film after annealing shows a more discrete LEED pattern (Figure 4f) and band structure (Figure 4h), which suggests less rotational disorder.

**4. Ge monolayer thickness determination via STM**

We determined the thickness of a Ge film monolayer from preliminary STM measurements of a thick Germanium film (nominal thickness 3.3 Å) evaporated on Au(111), the topography of which is shown in Figure S5. The STM image shows a number of Ge islands that allowed us to determine a Ge monolayer thickness of ~2.5 Å.

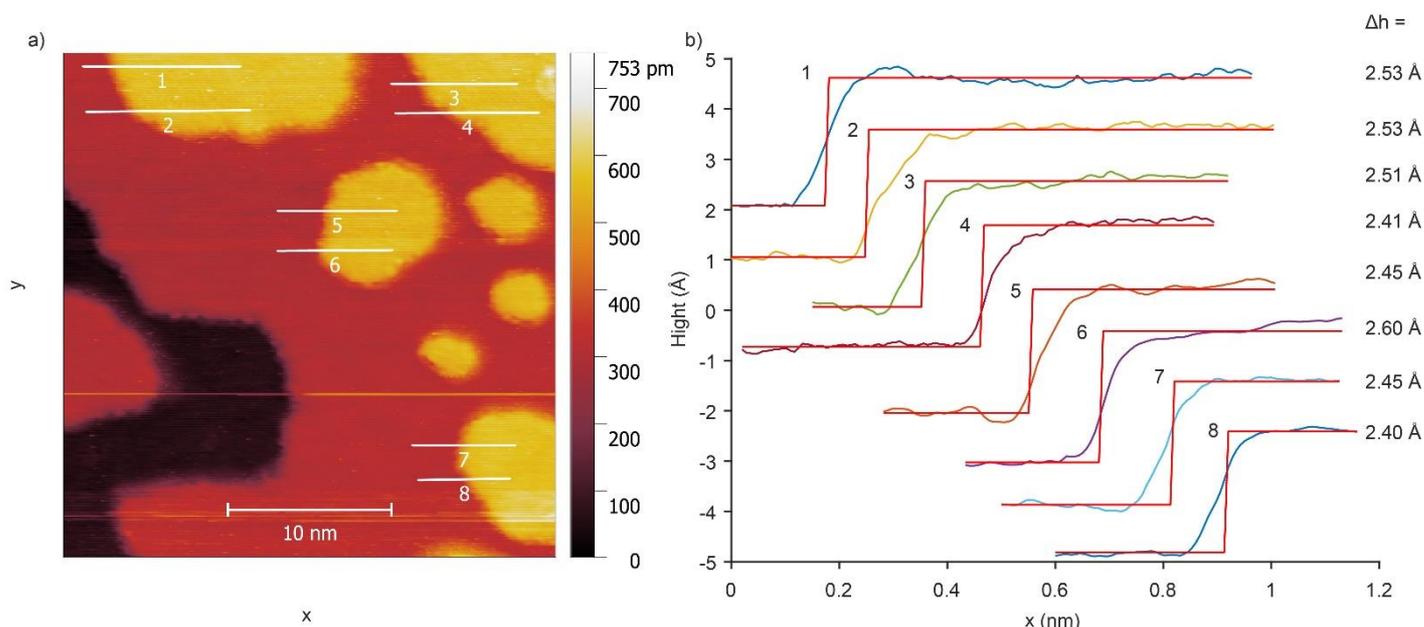

**Figure S5: Scanning tunneling microscopy measurements of Ge islands on Au(111).** a) Topography of Ge film with nominal thickness of 3.3 Å, showing the growth of Ge islands. b) Line-profile measurements corresponding to the white lines in a). Offsets were introduced to improve the visibility. The coloured curves indicate the measured data, while the red solid lines are the fitting curves from which the island heights Δh were extracted. The measurements were performed with a commercial Omicron STM at room temperature.

The structural transition that is indicated by the LEED measurements shown in Figure 1b-c is a strong indicator that at least one full layer of Germanium is completed at the nominally 2.6 Å thick films. which is also corroborated by the Ge appearance of a bulk (i.e. germanene) contribution in the Ge 3d core level peaks (shown in Fig 1a and S2). Due to the possibility of island growth, the film may be slightly thicker than one monolayer in some parts of the sample, but it is unlikely that two layers were completed in this film, since the interface component in the Ge 3d core level spectrum is still clearly present in Fig 1a for the nominally 2.6 Å thick film, but only fully suppressed for a ~2 ML thick film (shown in Fig SM2). Based on the STM measurements, quartz microbalance rate calibration, and core-level spectroscopy, we therefore conclude that the film with nominal thickness of 2.6 Å measured during our ARPES experiments is approximately at the monolayer limit.

**5. Influence of rotational disorder on the appearance of Dirac cones**

We observed the appearance of parallel lines on the equal energy surfaces shown in Figure 4(d). These lines can be understood from the suppositions of circular pockets with small rotational disorder, which will wash out the sides of the pocket, while leaving streaks of intensity tangential to the sense of the rotational disorder, as is illustrated by the simulation in Figure S6.

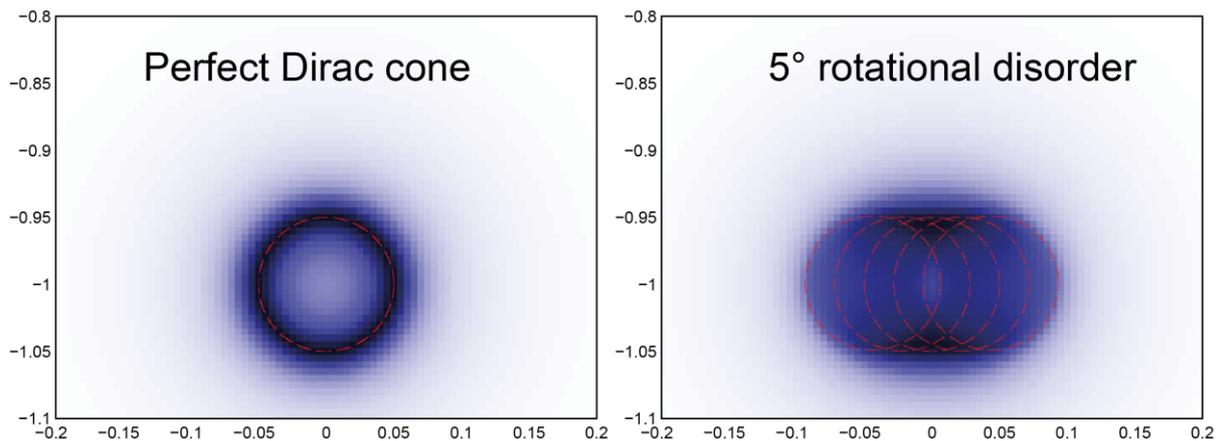

**Figure S6: Simulations of the effect of rotational disorder on the observations of Dirac cones.** Here the "perfect" cone is centered at (-1,0), while in the right panel we sum intensities after continuously rotating the cone over -2.5° to 2.5° (the centre of rotation is (0 0)).